# Computational investigation of 2D Phosphorene sheet towards NH$_3$ gas sensing

Naresh Kumar[1], Yogendra K. Gautam[1], Soni Mishra[2], Anuj Kumar[1*], Abhishek Kumar Mishra[3] *

[1] Department of Physics, Chaudhary Charan Singh University (CCS), Meerut, U.P., India.

[2] Department of Physics, Graphic Era Hill University, Dehradun 248002, India

[3] Applied Science Cluster, Department of Physics, School of Advanced Engineering, UPES University, Bidholi, Dehradun-248007, U.K., India.

**Abstract**

First-principles based calculations were executed to investigate the sensing properties of ammonia gas molecules on two-dimensional pristine black phosphorene towards its application as a gas sensor and related applications. We discuss in detail, the interaction of ammonia gas molecules on the phosphorene single sheet through the structural change analysis, electronic band gap, Bader charge transfer, and density-of-states calculations. Our calculations indicate that the phosphorene could be used as a detector of ammonia, where good sensitivity and very short recovery time at room temperature have confirmed the potential use of phosphorene in the detection of ammonia.

*Corresponding Author(s) email:

dranujkumarccsu@gmail.com (AK); mishra_lu@hotmail.com; akmishra@ddn.upes.ac.in (AKM)

# 1. Introduction

Gas detection is very important in industries and their peripheral monitoring and gas detection technology is expanding gradually with the use of high precision, low power and the cost-effective gas sensors. Oil and gas refineries, industrial factories, agriculture, environmental pollution policy, public safety, and the medical field are some of the major fields for the applications of gas sensors [1]. The large surface-to-volume ratio is an important factor in the field of gas detection technology. Due to the low noise level and high electrical conductivity, a small change in the concentration of carrier generates an imprudent change in electric conductivity upon gas exposure. Gas detection technology can use the change in electric conductivity as a determining factor [2-3].

Ammonia ($NH_3$) is a colorless gas used in wastewater treatment, leather, rubber, paper, fertilizers, biofuels, textiles, and food & beverage industries [4]. It is also used in cold storage, refrigeration systems, production of pharmaceuticals, printing, cosmetics industries, and fermentation [5]. However, this highly noxious, irritating, and suffocating gas is dangerous to public health and also hurts the surrounding environment [6]. Hence, high-quality gas sensors for $NH_3$ detection have been developed and monitored widely in recent years [7]. It is important to detect the low concentration of ammonia gas molecules in the industry with an efficient and appropriate technique. Therefore, many materials and techniques have been explored to detect this gas [7-9]. The high mobility, tunable band gap, and excellent transport properties of two-dimensional (2D) materials have motivated researchers to investigate novel devices based on 2D materials for gas sensing applications. Graphene, 2D transition metal dichalcogenides (TMDs), $MoS_2$, mxenes, silicene, borocarbide, and borophene, etc. are some of the examples of materials that have been widely studied and have the potential to develop advanced gas sensing devices [10-15].

Black phosphorene is the recently examined 2D material that has applications in the field of gas sensing. Black phosphorene (BP) is a layered form of black phosphorous atoms that exhibits a limited band gap and a large charge carrier mobility of about 1000 $Cm^2V^{-1}s^{-1}$ [16]. Phosphorene was first introduced in 2014 as a strong competitor to graphene which was invented in 2004 [17]. Unlike semi-metallic graphene, pristine phosphorene is a semiconductor with high carrier mobility and a finite band gap at room temperature, making it a promising candidate for nanoelectronics and gas sensing applications. Black phosphorene is used in field effect transistors, batteries, thin film solar cells, photovoltaic, and gas sensing applications [18-21]. Phosphorene exhibits an anisotropic electric conductivity, high current on/off ratio, and high-frequency execution when used as channels in the transistors [22]. Figure-1 shows the unit cell of black phosphorene. The lattice structure of black phosphorene is a hexagonal, honeycomb with a stiff and puckered shape. The band gap of the black phosphorene has a value changeable from 0.3 to 2 eV [23] depending on sheet thickness. This reveals that the inter-layer Van der Waals interaction has a pivotal effect on the band gap. Taking the above-discussed properties and the high surface-to-volume ratio of black phosphorene, BP has the potential for gas detection applications.

In this study, the investigation of the interaction of $NH_3$ gas molecule on pristine phosphorene is investigated employing first-principles-based density functional theory (DFT) simulations. We explore the gas sensing mechanism through structural changes, interaction energy ($E_{int}$), charge transfer, the electronic density of states (DOS), and band structure calculations. We also report the sensitivity, recovery time, and sensor efficiency parameters of the phosphorene-based ammonia sensor.

## 2. Computational Methods

All the calculations were performed with the use of Quantum ESPRESSO (opEn Source Package for Research in Electronic Structure, Simulation, and Optimization) [24] package based on density-functional-theory (DFT). We used the generalized gradient approximation (GGA) and the Perdew-Burke-Ernzerhof (PBE) [25] functional with Projector-Augmented-Wave (PAW) pseudopotential for performing all the calculations. The Grimme-D2 method [26] is used to include the Van der Waals (VdW) correction for considering the long-range interaction.

We first built the phosphorene unit cell using structures available in the literature [27-29] and then a 3×3×1 supercell was established to investigate the $NH_3$ gas molecule interaction. To avoid the interactions between the periodic images the vacuum of 20 Å was introduced in z-direction. The geometry optimizations were carried out by taking the kinetic energy cut-off of wave function (ecutwfc) at 60 Ry and the kinetic energy cut-off of charge density (ecutrho) at 240 Ry. The charge density convergence standard of the self-consistent field (SCF) was $10^{-6}$ eV. Brillouin zone integrations are sampled with a 10 x 8 x 1 Monkhorst-Pack k-point grid [30] for the phosphorene structure and electronic band gap calculations. Supercell of 3x3 was modeled using a 5 x 4 x 1 Monkhorst-Pack k-point grid [30].

The energetic data and band gap structure are also calculated using the hybrid functional Heyd-Scuseria-Ernzerhof (HSE) [31-32]. For this purpose, we used 2×2×1 supercell due to our computational limitations.

The interaction energy ($E_{int}$) is the energy absorbed or released during the interaction between the NH$_3$ gas molecule and phosphorene sheet and was calculated using

$$E_{int} = E_{phosphorene-NH_3} - \left(E_{phosphorene} + E_{NH_3}\right) \quad\quad (1)$$

Where $E_{phosphorene-NH_3}$, $E_{phosphorene}$ and $E_{NH_3}$ represent the total energies of the NH$_3$ molecule interacting on the phosphorene sheet, phosphorene sheet, and isolated NH$_3$ molecule, respectively. $E_{int} < 0$ indicates that the interaction process is exothermic and gas interaction occurs simultaneously.

The sensing quality of a sensor can be determined in terms of gas response or sensitivity (S, %), which is often, described both experimentally and theoretically by Equation-2 [33].

$$S = \left|\frac{\sigma_{phosphorene-NH_3} - \sigma_{phosphorene}}{\sigma_{phosphorene}}\right| \times 100 \quad\quad (2)$$

In Equation-2 $\sigma_{phosphorene-NH_3}$ shows the electrical conductivity of the operating sensor upon NH$_3$ gas adsorption and $\sigma_{phosphorene}$ is the electrical conductivity of pristine phosphorene before gas adsorption.

The electrical conductivity is related to the band gap ($E_g$) and can be described by Equation-3.

$$\sigma = AT^{3/2} \exp\left[-\frac{E_g}{2k_B T}\right] \quad\quad (3)$$

Here the operating temperature ($T$) is in the unit of Kelvin, the Boltzmann constant $k_B$ is in eV/K, and A represents the proportionality constant (electrons/m$^3$/K$^{3/2}$) [34]. A depends on the system.

The sensor reusability is an important parameter to evaluate the performance of the sensor. For practical applications, the short recovery time i.e. fast regeneration is an important factor in the successful cyclic operation of a sensor. The recovery time is calculated by applying the conventional transition state theory (TST) using Equation-4 [34-35].

$$\tau = \frac{1}{\nu_0} \exp\left[-\frac{E_{int}}{k_B T}\right]$$ ---------------------------------------------------------- (4)

Here $\nu_0$ is the attempt frequency in $s^{-1}$(Hz), and $E_{int}$ represents the interaction energy (eV), which is calculated from Equation-1.

Surface interaction of $NH_3$ gas molecules on pristine phosphorene may give rise to a change in work function. The difference between the vacuum level and the vacuum level is termed as work function. The work function is calculated using Equation-5.

$$\varphi = E_{vac} - E_f$$ ---------------------------------------------------------- (5)

Where $E_{vac}$ is the vacuum energy level and $E_f$ is the Fermi energy of the system.

The sensor efficiency is also calculated from the relative change in the work function of the sensor using Equation-6 [34-35].

$$\varepsilon = 100\left[\left|\frac{\varphi_{phosphorene-NH_3}}{\varphi_{phosphorene}} - 1\right|\right]$$ ---------------------------------------------------------- (6)

Where $\varphi_{phosphorene-NH_3}$ and $\varphi_{phosphorene}$ are the work functions of phosphorene after adsorption of $NH_3$ gas molecule and before adsorption, respectively.

Bader charges [36-37] of the system have been analyzed to find out the electron transport between the $NH_3$ molecule and pristine phosphorene. The charge transfer quantity ($Q_t$) represents the change of electronic properties on the surface of phosphorene upon gas molecule interaction. $Q_t$ is defined as the charge transferred from a gas molecule to a phosphorene sheet in the process of gas interaction. The charge transfer is calculated using Equation-7 [38].

$$Q_t = Q_{absorbed(gas)} - Q_{isolated(gas)}$$ ---------------------------------------------------------- (7)

Here, $Q_{absorbed(gas)}$ and $Q_{isolated(gas)}$ are the carried charge of a gas molecule after adsorption and isolated gas molecule respectively.

## 3. Results and discussion

### 3.1 Structural model of Pristine Phosphorene and $NH_3$ gas molecule

As depicted in Figure-2, the phosphorene comprises two atomic layers of phosphorus atoms. These layers are ordered in a puckered honeycomb lattice. Different from hexagonal graphene, phosphorene has an orthorhombic unit cell that shows armchair and zigzag configuration along the x and y direction respectively, resulting in a remarkable structural anisotropy. Due to the sp$^3$ hybridization, phosphorene is not an atomically flat sheet-like graphene that has sp$^2$ hybridization. Instead, phosphorene contains puckered honeycomb structured layers [39]. Weak Van der Waals forces act between these layers, which are responsible for holding these layers together.

In our simulation model of pristine phosphorene, we first optimized the pristine phosphorene unit cell with 4 atoms as shown in Figure 1. The lattice constants of monolayer phosphorene are optimized to be 4.62 Å and 3.30 Å along armchair and zigzag directions respectively, which is in good agreement with the previous reports [40-42] as shown in Table-1.

**Table:-1** Lattice constants (a & b) and P-P bond length ($d_1$ & $d_2$) along armchair and zigzag direction respectively.

| Author | a (Å) | b (Å) | $d_1$ (Å) | $d_2$ (Å) |
|---|---|---|---|---|
| This work | 4.62 | 3.30 | 2.259 | 2.223 |
| Motohiko Ezawa [33] | 4.43 | 3.27 | 2.207 | 2.164 |
| J. Prasongkita et al. [34] | 4.63 | 3.31 | 2.280 | 2.250 |
| Ankit Jain et al. [35] | 4.43 | 3.28 | --- | --- |

In the optimized structure of phosphorene, the bond length of the P-P bond is $d_1$ (interlayer distance) = 2.259 Å in the armchair direction and $d_2$ (intralayer distance) = 2.223 Å in the zigzag direction. The values of $d_1$ and $d_2$ are slightly different but very close to each other due to the covalent bonds between phosphorus 3p orbitals. For comparison, Ezawa M [40] found the value of P-P bond length $d_1$ = 2.207 Å and $d_2$ = 2.164 Å in the armchair and zigzag directions respectively, and $d_1$ = 2.28 Å and $d_2$ = 2.25 Å were reported by Prasongkita J *et al.* [41] and our findings are in good agreement with these studies performed with DFT GGA functionals.

The calculated cohesive energy is -5.229 eV as listed in Table 2, whereas the work function of phosphorene is found to be 4.567 eV which is nearly equal to the value (4.50 eV) obtained by Cai Y et al. [43] using DFT with PBE functional. We also investigated the magnetic nature of phosphorene and found the magnetic moment to be 0.0 Bohr magneton which confirms the non-magnetic nature of pristine phosphorene 2D sheet. The non–magnetic nature of phosphorene was also obtained by Kim S et al. [44].

To further analyze the electronic properties of phosphorene, the density of states and band structure were plotted using GGA-PBE and HSE hybrid functionals. Figure 3 shows the DOS and band structure of pristine phosphorene with GGA-PBE functional. The band structure (Figure-3(b)) reveals the semiconductor character of phosphorene.

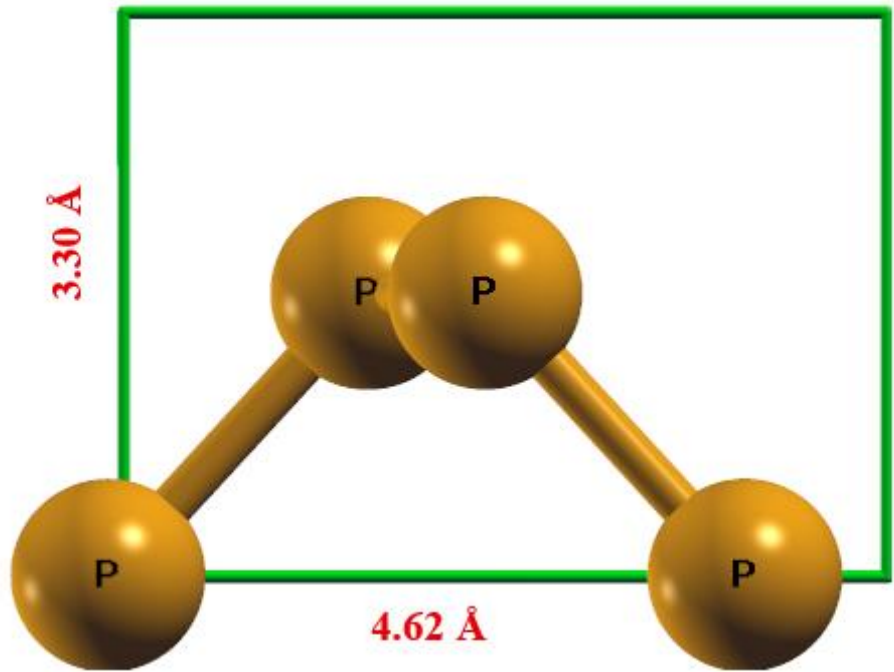

**Figure: 1** The unit cell of pristine phosphorene.

**Table. 2** Cohesive energy of phosphorene, adsorption energy of P-NH$_3$, band gap (GGA Functional) of phosphorene & P-NH$_3$, and charge transfer during NH$_3$ adsorption.

| Structure | $E_{coh}$ (eV) | $E_{int}$ (eV) | $E_g$ (eV) | $\varphi(eV)$ | $Q_t$ (e$^-$) |
|---|---|---|---|---|---|
| **Phosphorene** | -5.229 | --- | 0.902 | 4.567 | --- |
| **Phosphorene-NH$_3$** | --- | -0.054 | 0.914 | --- | -0.013 |

Phosphorene has the direct band gap at the Γ point of the Brillouin zone and it is calculated as 0.902 eV using GGA-PBE as shown in Table 2. This value is near the value of the band gap 0.84 eV calculated by Cai Y et al. at the GGA level [43]. A. S. Rodin et al. [45] also calculated the band gap of phosphorene at the GGA level with a value of 0.7 eV using the PBEsol functional.

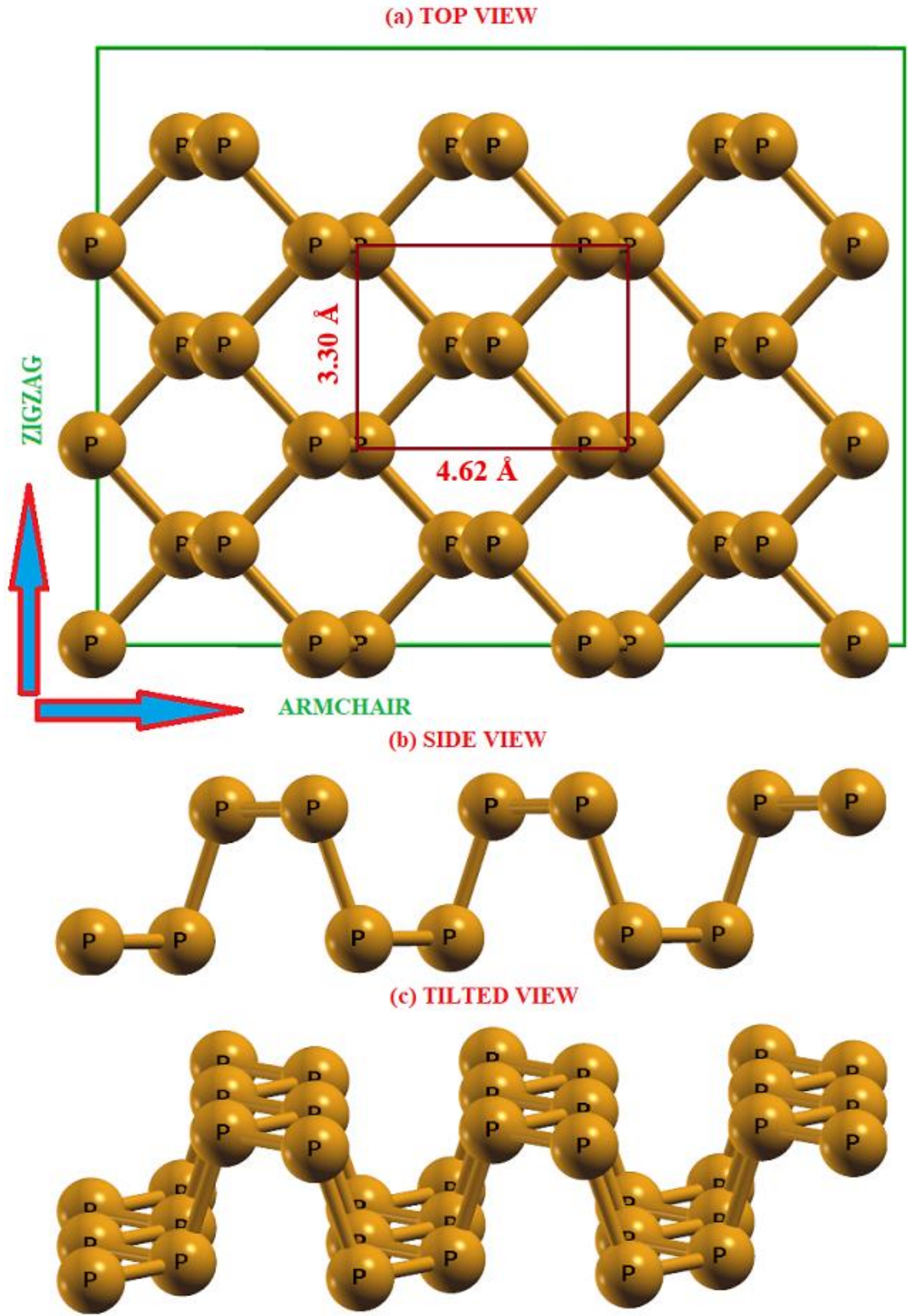

**Figure: 2** DFT Calculated relaxed structure of pristine phosphorene (a) Top view, (b) Front view, and (c) Tilted view.

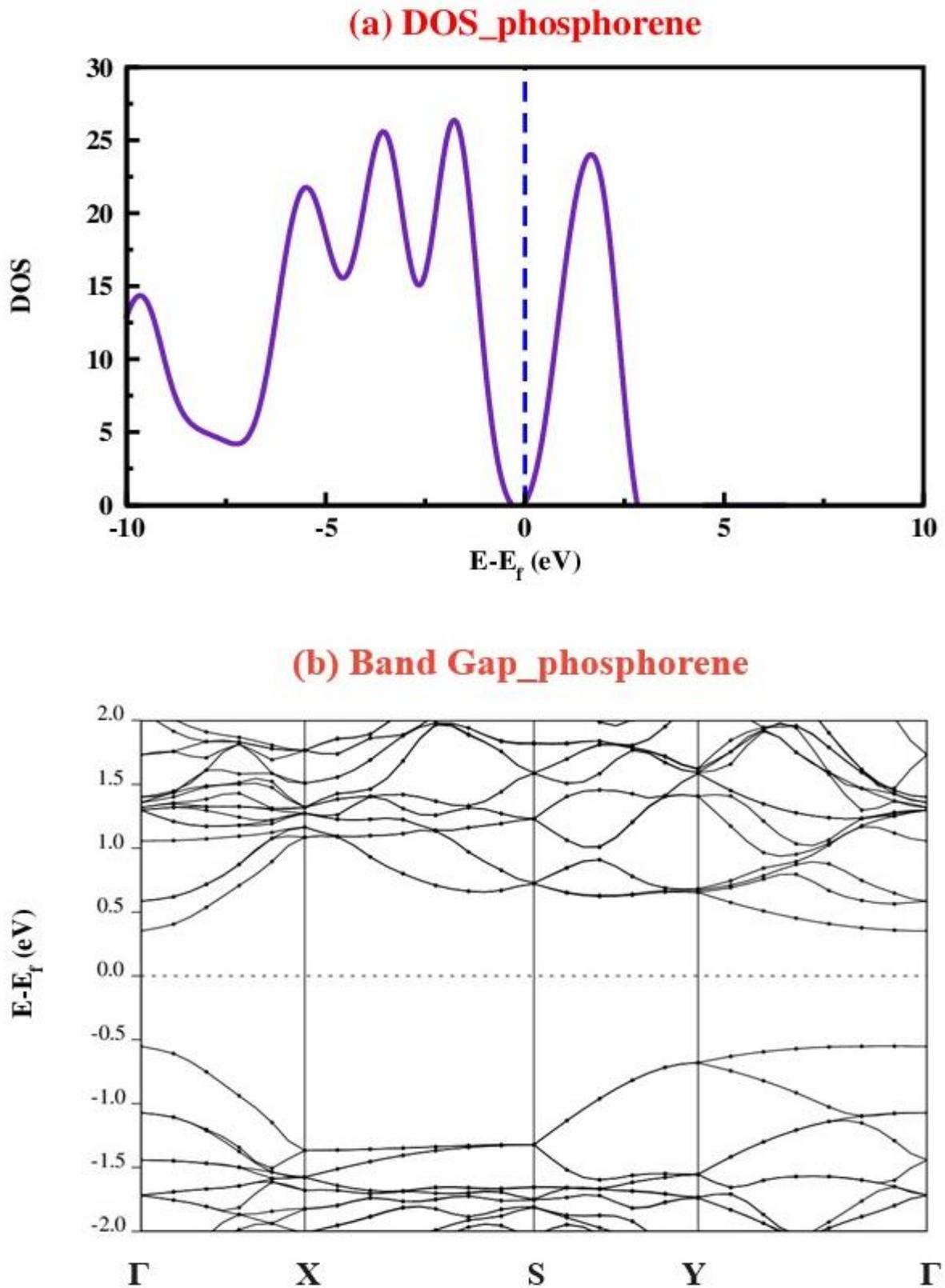

**Figure: 3 (a)** Density of states and **(b)** Band structure of phosphorene calculated using GGA-

PBE functional.

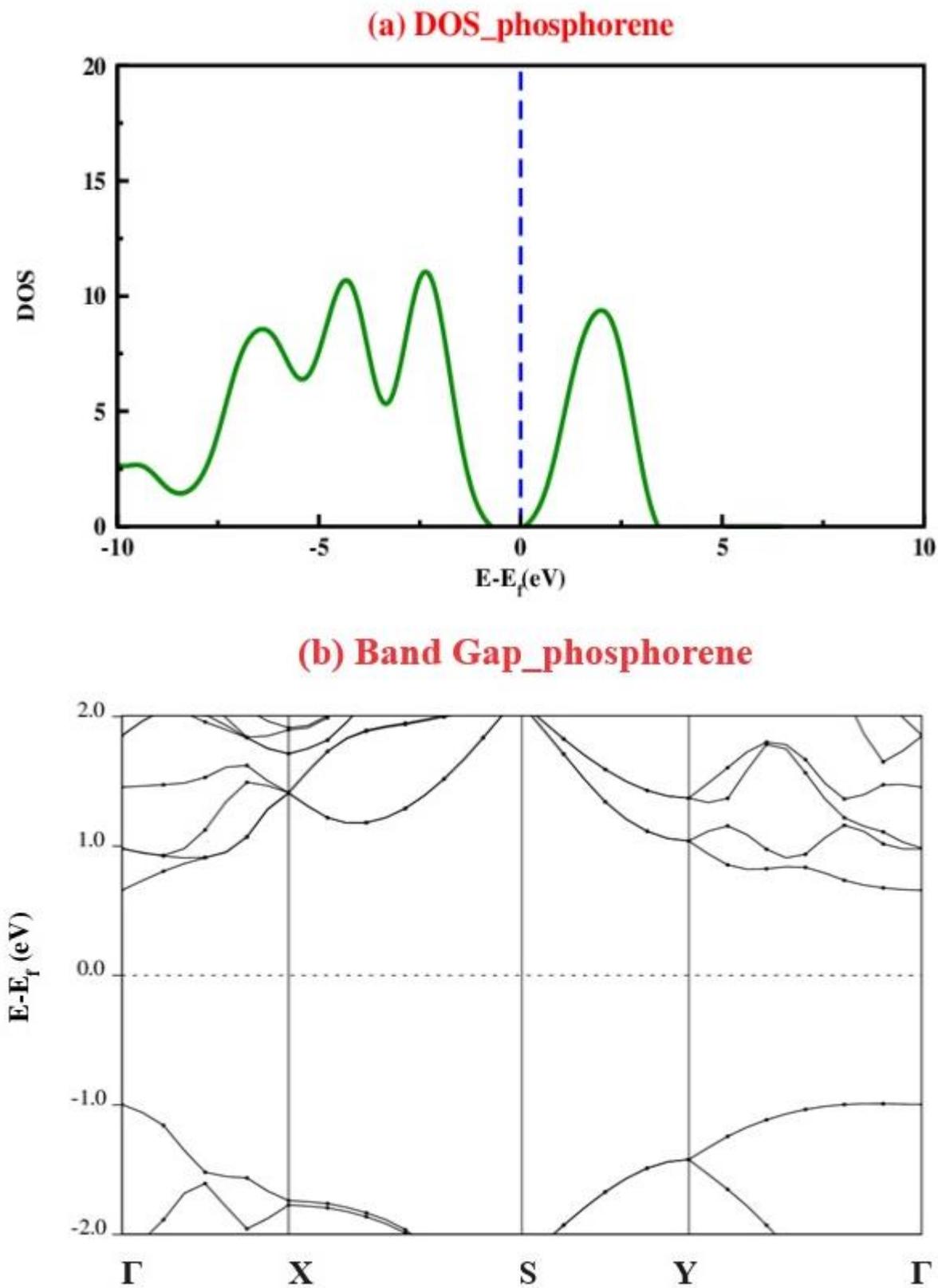

**Figure: 4 (a)** Density of states and **(b)** Band structure of phosphorene calculated using HSE

hybrid functional.

We have given density -of -states (DOS) of phosphorene sheet, calculated using GGA-PBE level, in (Figure 3(a)). We observe that DOS is maximum in valance band, while there is only one DOS peak in conduction band.

Figure 4 shows DOS and band structure of phosphorene calculated using HSE hybrid functional. According to HSE calculations, band gap of phosphorene is 1.64 eV which is close to the earlier work of Cai Y et al. [43] who found a band gap of 1.52 eV at the HSE level. DOS diagram at the HSE level shows the same result as we obtain at the GGA level except higher band gap. The magnitude of DOS is less in HSE calculation because we consider 2×2×1 supercell in HSE calculations.

**3.2 Interaction of Ammonia (NH$_3$) on pristine phosphorene**

To investigate the possible interaction between an ammonia molecule and a phosphorene sheet, an ammonia molecule was placed close to phosphorene in different directions and configurations and found that the NH$_3$ molecule interacts with the phosphorene through an interaction energy of -0.054 eV (Table 2). This low value of interaction energy indicates that the adsorption of NH$_3$ on pristine phosphorene is physical adsorption. The interaction process is simultaneous and exothermic. In the case of ammonia adsorption on graphene, the adsorption energy is -0.029 eV which was calculated by Zhou M et al. [46] using DFT with GGA-PBE functional. Our calculated E$_{int}$ value is better than this value and adsorption of NH$_3$ on pristine phosphorene is more stable. Hence phosphorene is a more promising material for ammonia gas sensing. Here, the distance between ammonia gas molecules to the phosphorene sheet is found to be 3.39 Å. The optimized structure of the P-NH$_3$ system is shown in Figure 5.

We next performed the Bader charge analysis to further quantify this interaction and found the charge transfer to be -0.013e$^-$ as listed in Table 2. The negative value of Q$_t$ indicates the electrons get transferred from pristine phosphorene to NH$_3$ molecule, and NH$_3$ shows the characteristics of the electron acceptor.

The band structure of the phosphorene-ammonia (P-NH$_3$) system is shown in Figure 6 (b), which is calculated at GGA-PBE level. Upon interaction with the NH$_3$ molecule the semiconductor nature of pristine phosphorene becomes stronger as after the interaction the band gap is enhanced to 0.926 eV as a result of the transfer of electrons from to pristine phosphorene to the $NH_3$ molecule. The DOS of the P-$NH_3$ system is depicted in Figure-6(a).

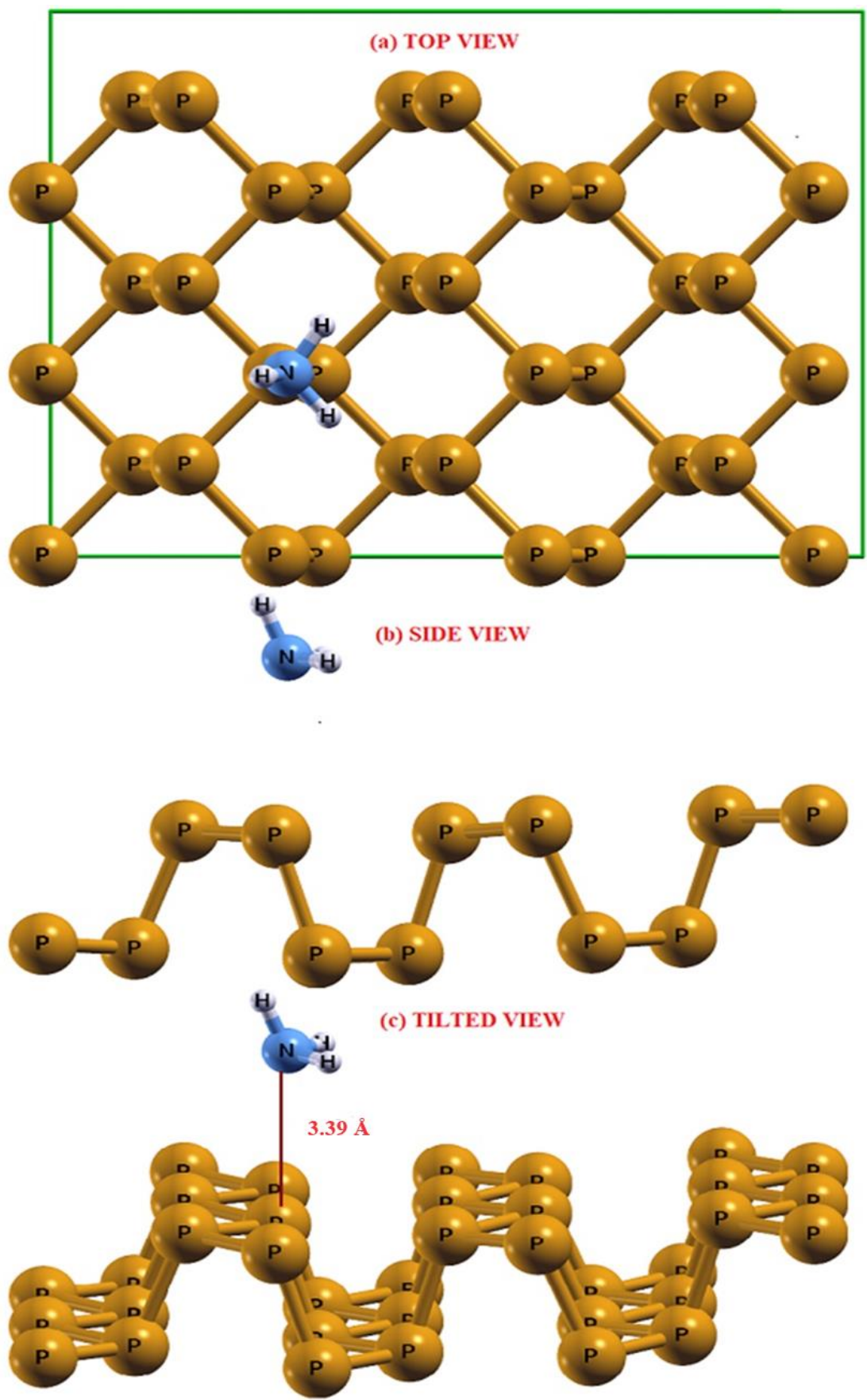

**Figure: 5** Optimized structures of P-NH$_3$ **(a)** Top view, **(b)** Side view, and **(c)** Tilted view.

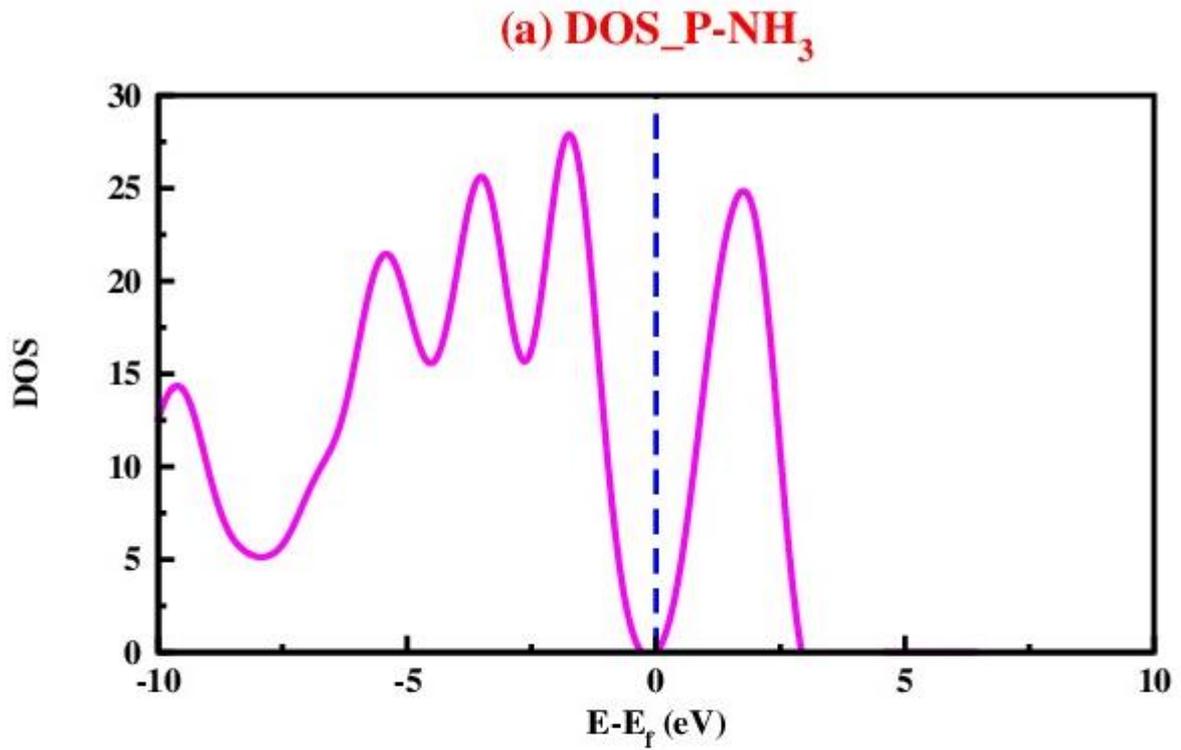

**Figure:6 (a)** Density of states and **(b)** Band structure of phosphorene after NH₃ adsorption calculated at GGA-PBE level.

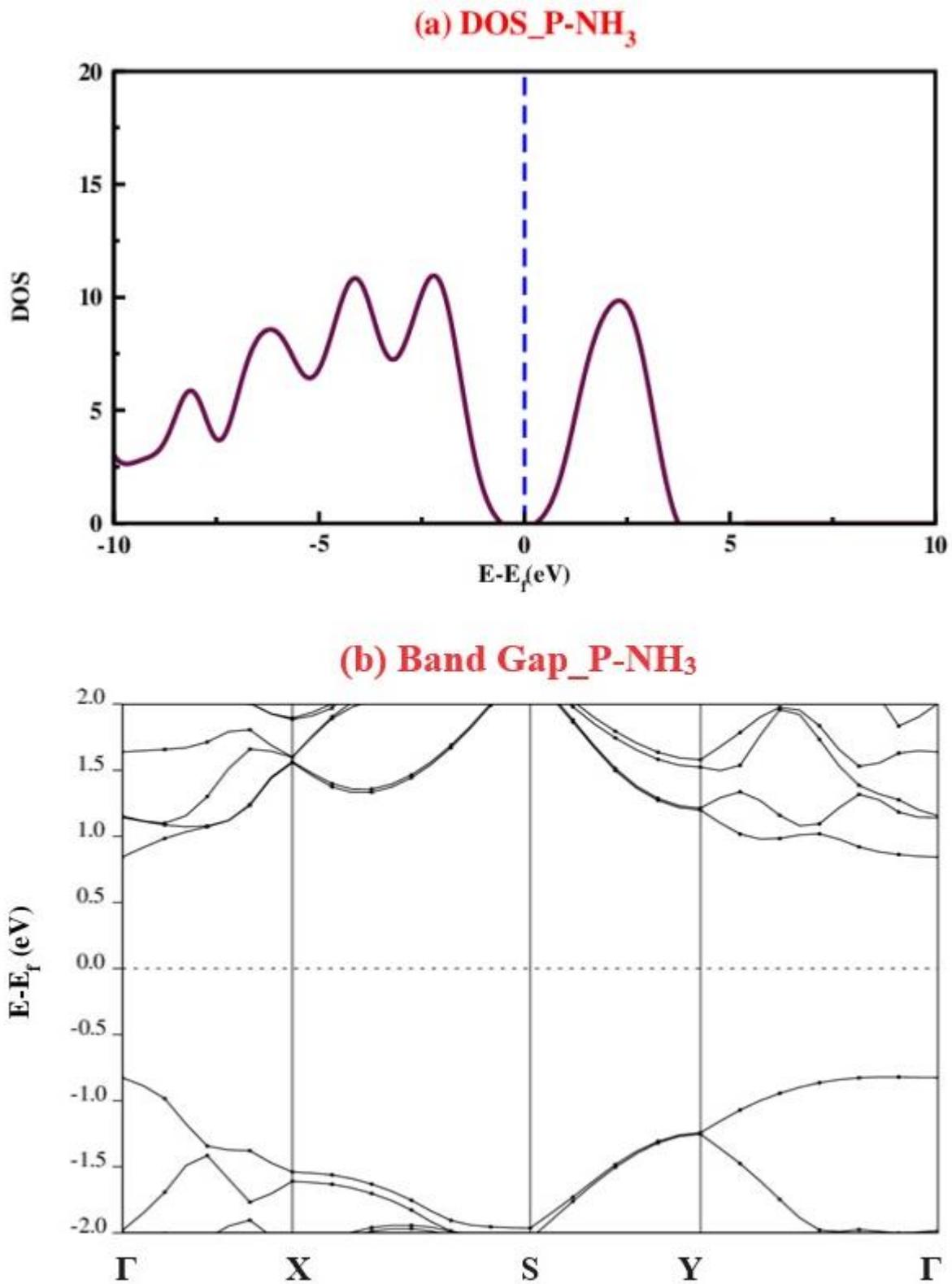

**Figure: 7 (a)** Density of states and **(b)** Band structure of phosphorene after $NH_3$ adsorption calculated using HSE functional.

We also performed calculations using HSE functionals and the band gap of phosphorene increases from 1.64 eV to 1.66 eV upon $NH_3$ adsorption. The DOS and band structure of P-$NH_3$ system at HSE level are shown in Figure-7(a) and 7(b) respectively.

In order to investigate the effect of functionals on our calculated interaction energy values, we compared calculated interaction energies by GGA and HSE functionals. We did calculations on a 2x2x1 supercell and found interaction energies to be 0.045 and 0.042 respectively at GGA and HSE functionals showing no signification changes.

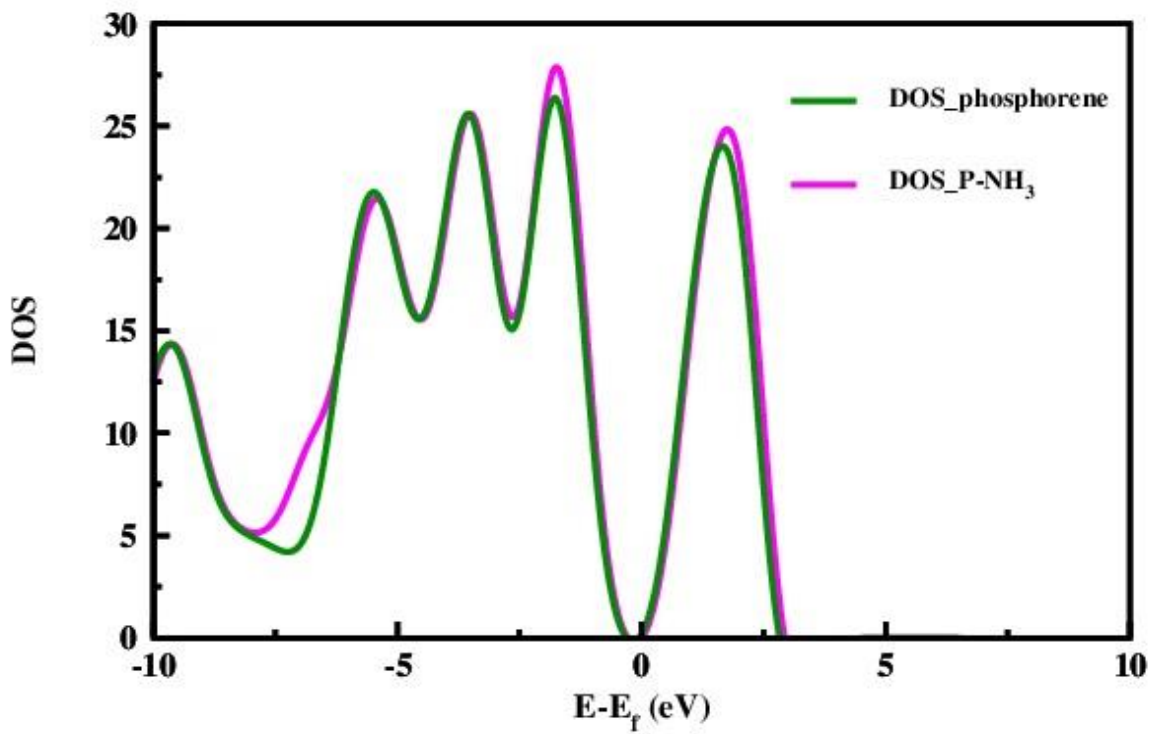

**Figure: 8** Comparison graph of DOS of phosphorene and P-$NH_3$

The Figure-8 shows the comparison of DOS of phosphorene before and after adsorption of $NH_3$ and we note a slight shift in the states towards higher energy side as a result of interaction.

We list few key performance parameters related to the pristine phosphorene to sense the $NH_3$ gas molecule in Table-3. The sensitivity to the ammonia gas on the pristine phosphorene is calculated to be 20.06 % at room temperature. For comparison the sensitivity of pristine phosphorene for phosgene gas was found to be 16 % by Ghambarian M. et al. [35]. Ghadiri M. et al. [34] calculated sensitivity for hydrogen sulphide adsorbed on phosphorene to be 1.6 % only. Luo H et al. [47] calculated the sensitivity of intrinsic graphene for $NH_3$ gas and they found the value below 1 %. Our calculated high value of sensitivity of pristine phosphorene for the adsorption of ammonia indicates that phosphorene can become a good material to detect the ammonia gas molecule. Table 3 also shows the recovery time of pristine phosphorene to detect ammonia. A very short recovery time of the value 7.90 picosecond (ps) at room temperature indicates that pristine phosphorene is a very fast recoverable sensor toward $NH_3$ at visible frequency. For comparison, Ghadiri M et al. [34] found the value of recovery time of 0.1 ps for the detection of hydrogen sulphide on phosphorene. The detection efficiency of the pristine phosphorene sensor for $NH_3$ gas is about 3 %.

**Table:-3** Sensitivity, recovery time, and sensor efficiency of phosphorene-$NH_3$.

| Structure | S (%) | $\tau$ (ps) | $\varepsilon$ (%) |
|---|---|---|---|
| **Phosphorene-$NH_3$** | 20.06 | 7.90 | 3.36 |

## 4. Conclusion

In this study, we have investigated the $NH_3$ gas molecule sensing properties of pristine phosphorene sheet using density function theory calculations. We have done an investigation of $NH_3$ interaction with pristine phosphorene by calculating interaction energy, charge transfer, band structure, and DOS before and after the interaction of gas molecules. In order to check the sensor quality we have calculated the sensitivity, recovery time, and sensor efficiency. Good sensitivity and very short recovery time confirm the potential use of phosphorene in the detection of ammonia. In our study, we find that pristine phosphorene has the potential to sense the ammonia gas in the environment.

**Acknowledgments**

NK is grateful to the Principal, D. N. (P.G.) College Gulaothi, Bulandshahr for allowing him to pursue a Ph.D. AKM acknowledges SERB SURE (SUR/2022/004935) grant (year 2023).